\documentclass{cernyrep}
\usepackage[T1]{fontenc}
\usepackage[bookmarks, colorlinks=true, linktoc=page, pdftex, linkcolor=black, citecolor=black, urlcolor=blue]{hyperref}
\usepackage{fancyhdr}
\fancyhfoffset{4 mm}
\fancypagestyle{ARTICLE}{%
\fancyhf{} 
\lhead{\hfill CERN Accelerator School Proceedings ---
{\it RF for Accelerators} ---  Berlin, Germany,  2023\hfill}
\lfoot{\hspace{3 mm} Available online at \url{https://cas.web.cern.ch/previous-schools}}
\rfoot{\thepage\hspace*{3mm}}

}
\fancypagestyle{ARTTITLE}{%
\fancyhf{}
\lhead{\hfill CERN Accelerator School Proceedings ---
{\it RF for Accelerators} ---  Berlin, Germany,  2023\hfill}
\lfoot{\hspace{3 mm} Available online at \url{https://cas.web.cern.ch/previous-schools}}
\rfoot{\thepage\hspace*{3mm}}

}

\usepackage{varwidth}
\usepackage{xcolor}
\usepackage{graphicx}
\usepackage{epsfig}
\usepackage{bm}
\usepackage{amsmath}
\usepackage{longtable}
\usepackage{mathtools}
\usepackage{siunitx}
\usepackage{appendix}
\usepackage{todonotes}
\newcommand{\E}{\mathcal{E}}

\begin{document}

\title{Longitudinal beam dynamics II. Lecture 1: single and double rf systems}
\author{Elena Shaposhnikova\thanks{Elena.Chapochnikova@cern.ch}}

\date{\today}
\begin{abstract}

In the first part of this lecture, the main characteristics of synchrotron motion,  namely, synchronous particle and phase, bucket area and emittance, are introduced. The equations of longitudinal particle motion in a single rf system of a~proton synchrotron are re-derived and the phase equation is analyzed  for small and large amplitude particles. The particle motion below and above transition energy is also considered.
The main applications and operation modes of a double rf system are discussed in the second part of the lecture.
The intensity effects and beam instabilities are the subject of the following lecture.
\end{abstract}

\maketitle
\thispagestyle{ARTTITLE}

\section{\label{sec:Introduction}Introduction}
\noindent
This lecture is a continuation of the lecture {\it Longitudinal beam dynamics I} by F.~Tecker at this School, where the basic concepts of the longitudinal motion in a synchrotron were introduced (see also the~textbook~\cite{Lee}). Here the equations of longitudinal motion are re-derived in the form and notations which are used in the following lecture on intensity effects and beam instabilities, and in the papers~\cite{Ivan, Karpov2024} on which this second lecture is based.

The lecture is organized in the following way. After this {\bf Introduction}, the main equations of longitudinal motion in a single rf system without intensity effects are derived in {\bf Section~\ref{sec:main_equations}}. The phase equation is analyzed for small and large amplitude particles, the rf bucket, bunch emittance and synchrotron frequency are defined, and the particle motion during acceleration, below and above transition energy, is considered.
In {\bf Section~\ref{sec:double_rf}}, the main applications and operation modes of a double rf system in hadron rings are discussed with examples from the CERN accelerators. Some of the results presented here are used in the next lecture, in particular, for the discussion of the beam instability cures.

\section{\label{sec:main_equations} Longitudinal motion
in a single RF system}
\subsection{Synchronous particle, equations of motion}
The longitudinal motion of particles is also often called synchrotron motion.
In a synchrotron, the average particle orbit is constant during acceleration, thus the magnetic field $B$ and the RF frequency $\omega_{RF}$ should increase synchronously. The design momentum $p_s$ follows the magnetic field variation
\[
p_s= q B \rho,
\]
where $q$ is the particle charge and $\rho$ is the bending radius.   The corresponding particle has a revolution period $T_0 = 2\pi R/\beta_0 c$ and angular revolution frequency $\omega_0=2\pi f_0=2\pi/T_0$, where $R$ is the average machine radius.

A particle synchronized with the RF frequency $\omega_{RF} = h \omega_0$ is called the synchronous particle; its energy gain per turn is
\begin{equation}
\delta E_0 = q V_0 \sin\phi_s,
\label{gain}
\end{equation}
where $\phi_s$ is the synchronous phase,  $h$ the harmonic number, and $V_0$ the RF voltage amplitude.
The~acceleration rate of the synchronous particle is
\begin{equation}
    \frac{\delta E_0}{T_0} = \frac{dE_s}{dt}=\frac{\omega_0}{2\pi}\,qV_0 \sin\phi_s,
    \label{eq:acc_rate}
\end{equation}
so that in this phase convention there is no acceleration if $\phi_s =0$ or $\pi$.
A non-synchronous particle
has small deviations $\Delta p$, $\Delta E$, $\Delta \omega$, and $\Delta \phi$ from the corresponding parameters $p_s$, $E_s$, $\omega_0$, and $\phi_s$ of the~synchronous particle.
The energy gain per turn for this particle is
$\delta E = q V_0 \sin\phi$,  and the acceleration rate
\begin{equation}
    \frac{\delta E}{T_0} = \frac{dE}{dt}=\frac{\omega_0}{2\pi}\,qV_0 \sin\phi,
\end{equation}
where $\phi= \phi_s +\Delta \phi$. Subtracting the similar equation (\ref{eq:acc_rate}) for the synchronous particle, we obtain the first equation of longitudinal motion
\begin{equation}
    \frac{d}{dt}\left(\frac{\Delta E}{\omega_0}\right) = \frac{1}{2\pi}\,qV_0 (\sin\phi- \sin\phi_s).
    \label{eq:l1}
\end{equation}
Consider now the phase of a particle relative to the RF voltage,  $\phi (t)= \int \omega_{RF}\,dt -h\theta(t)$, where $\theta = \int\omega dt$ is its azimuthal position, so that $\dot{\phi} = -h\Delta\omega$.
Using the relation
\begin{equation}
    \frac{\Delta\omega}{\omega_0} = - \eta \frac{\Delta p}{p},
\end{equation}
which is also the definition of the slip factor $\eta = 1/\gamma_{t}^2 - 1/\gamma^2$, we have
\begin{equation}
    \frac{d\phi}{dt} = - h\omega_0\frac{\Delta\omega}{\omega_0} = h\omega_0\eta \frac{\Delta p}{p}.
\end{equation}
Using the kinematic relation $\frac{\Delta p}{p} = \frac{1}{\beta^2} \frac{\Delta E}{E}$, one finally obtains the second equation of longitudinal motion
\begin{equation}
    \frac{d\phi}{dt}  = \frac{h\omega_0^2\eta}{\beta^2 E}\frac{\Delta E}{\omega_0}.
    \label{eq:l2}
\end{equation}

\subsection{Phase equation}
\subsubsection{Small amplitude oscillations}
Combining the two equations of motion in conjugate variables $\left(\frac{\Delta E}{h\omega_0},\phi\right)$, Eqs.~(\ref{eq:l1}) and (\ref{eq:l2}),  and assuming a slow time variation of the parameters $E_s$, $\omega_0$, and $\eta$, one obtains the phase equation
\begin{equation}
     \frac{d^2\phi}{dt^2} = \frac{h\omega_0^2\eta}{2\pi\beta^2 E}\,qV_0 (\sin\phi- \sin\phi_s),
    \label{eq:phase}
\end{equation}
which describes the synchrotron oscillations of a particle around the synchronous phase $\phi_s$.
For particles with small oscillation amplitudes, $\Delta \phi = \phi - \phi_s \ll 1$, it reduces to the well-known equation of a~harmonic oscillator
\begin{equation}
    \frac{d^2\Delta\phi}{dt^2}  + \omega_{s0}^2 \Delta\phi =0,
    \label{eq:lin_phase}
\end{equation}
with the frequency of linear synchrotron oscillations defined as
\begin{equation}
    \omega_{s0}^2 = - \frac{h\omega_0^2\eta \cos\phi_s}{2\pi\beta^2E} q V_0.
    \label{ws}
\end{equation}
This equation determines the phase stability of particles.
Indeed, the motion is bounded if $\omega_{s0}^2 >0$, which requires $\eta\cos\phi_s <0$. In the absence of acceleration,
this means that below transition energy, for $\gamma < \gamma_{t}$, the synchronous phase is $\phi_s=0$, and, correspondingly, it
should be $\pi$ for  $\gamma > \gamma_{t}$. The phase stability during acceleration is considered later.

The solutions of the linear phase equation Eq.~(\ref{eq:lin_phase}) are
\begin{align}
    \Delta\phi & = \Delta\phi_m\cos \psi,\\
    \Delta\dot{\phi} & = - \Delta\dot{\phi_m}\sin \psi,
\end{align}
where  $\psi =\omega_{s0} t +\psi_0$ is the phase of synchrotron oscillations, and
$\Delta\dot{\phi} = \frac{d\Delta\phi}{dt}$ is connected with
$\left(\frac{\Delta E}{h\omega_0}\right)$ via Eq.~(\ref{eq:l2}), so that
\[
\Delta\dot{\phi}_m = \omega_{s0}\Delta\phi_m = \frac{\omega_{RF}^2\eta}{\beta^2E} \left(\frac{\Delta E_m}{h\omega_0}\right).
\]

\subsubsection{Large amplitude particles}
The phase equation (\ref{eq:phase}) describes the motion of a nonlinear oscillator. Using the definition of the linear synchrotron frequency (\ref{ws}), it can be rewritten as
\begin{equation}
    \frac{d^2\phi}{dt^2} +\frac{\omega_{s0}^2}{\cos\phi_s}(\sin\phi- \sin\phi_s) =0.
    \label{eq:phase_s}
\end{equation}
Multiplying this equation by $\dot{\phi}$ and integrating over $t$, we obtain an integral of motion $\E$, which is also called the energy of synchrotron oscillations
\begin{equation}
    \frac{\dot{\phi}^2}{2\omega_{s0}^2} +U(\phi) = \E,
\end{equation}
where the RF potential $U(\phi)$ in the general case  is defined as
\begin{equation}
    U(\phi) = \frac{1}{V_0\cos\phi_s} \int_{\phi_s}
    ^{\phi} [V(\phi^\prime)- V_0\sin\phi_s] d\phi^\prime.
\end{equation}
For a single RF system with voltage $V(\phi) = V_0\sin{\phi}$, one has (omitting the constant term $-U(\phi_s)$)
\begin{equation}
    U(\phi) = - (\cos\phi +\phi \sin \phi_s)/\cos\phi_s.
    \label{eq:U}
\end{equation}

The trajectories of a particle in the longitudinal phase-space $(\dot{\phi}/\omega_{s0}, \phi)$ are described by the equation
\begin{equation}
    \dot{\phi} = \pm \omega_{s0} \sqrt{2[\E-U(\phi)]}.
    \label{traj}
\end{equation}
\subsubsection{Acceleration. RF bucket}
\begin{figure}[h]
\centering
\includegraphics[width = 0.6\textwidth]{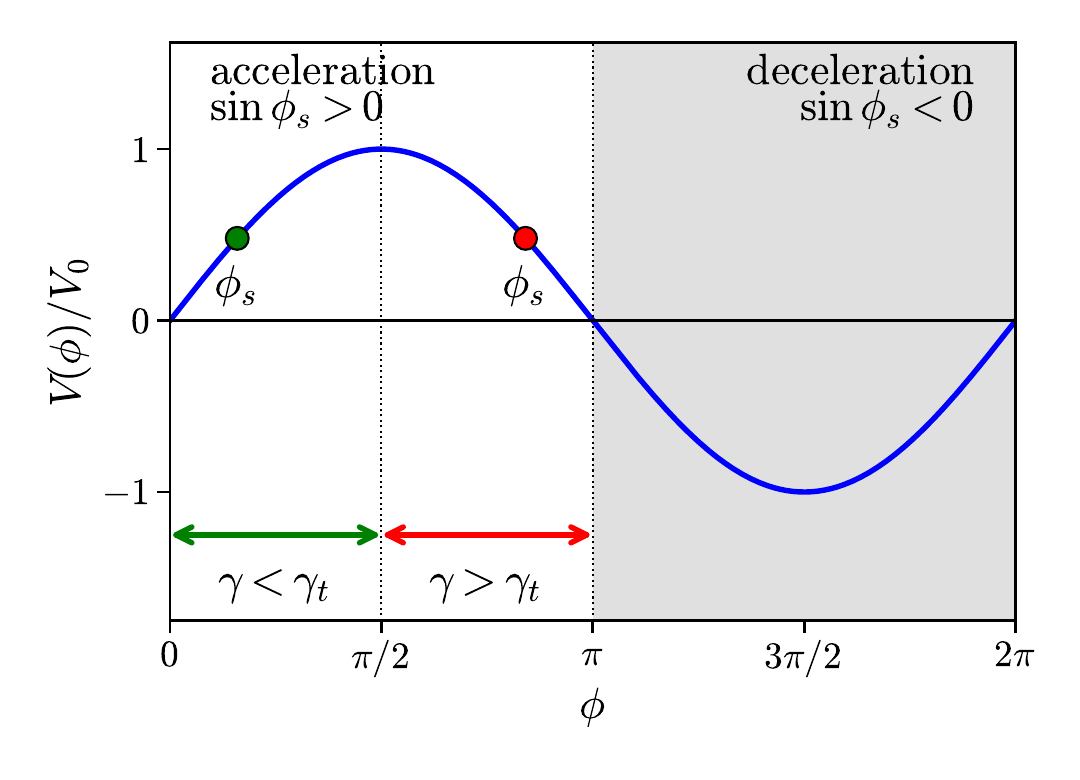}
\caption{Range of the synchronous phase $\phi_s$ required for beam acceleration, from the phase stability condition $\eta\cos\phi_s<0$: the green arrow shows the allowed range $0<\phi_s<\pi/2$ below transition energy ($\eta<0$) and the red arrow the range $\pi/2<\phi_s<\pi$ above it ($\eta>0$). Examples of $\phi_s$ below (green) and above (red) transition energy are shown by dots.}
\label{fig:phase_stab}
\end{figure}

The synchronous particle has a positive energy gain, see Eq.~(\ref{gain}), only if $\sin\phi_s >0$. Together with the~phase stability condition $\eta\cos\phi_s<0$ discussed above, this requirement determines (see Fig.~\ref{fig:phase_stab}) the~possible values of the synchronous phase $\phi_s$ below and above transition energy, which do not overlap.

\begin{figure}[h]
	\centering
 \includegraphics[width = \textwidth]{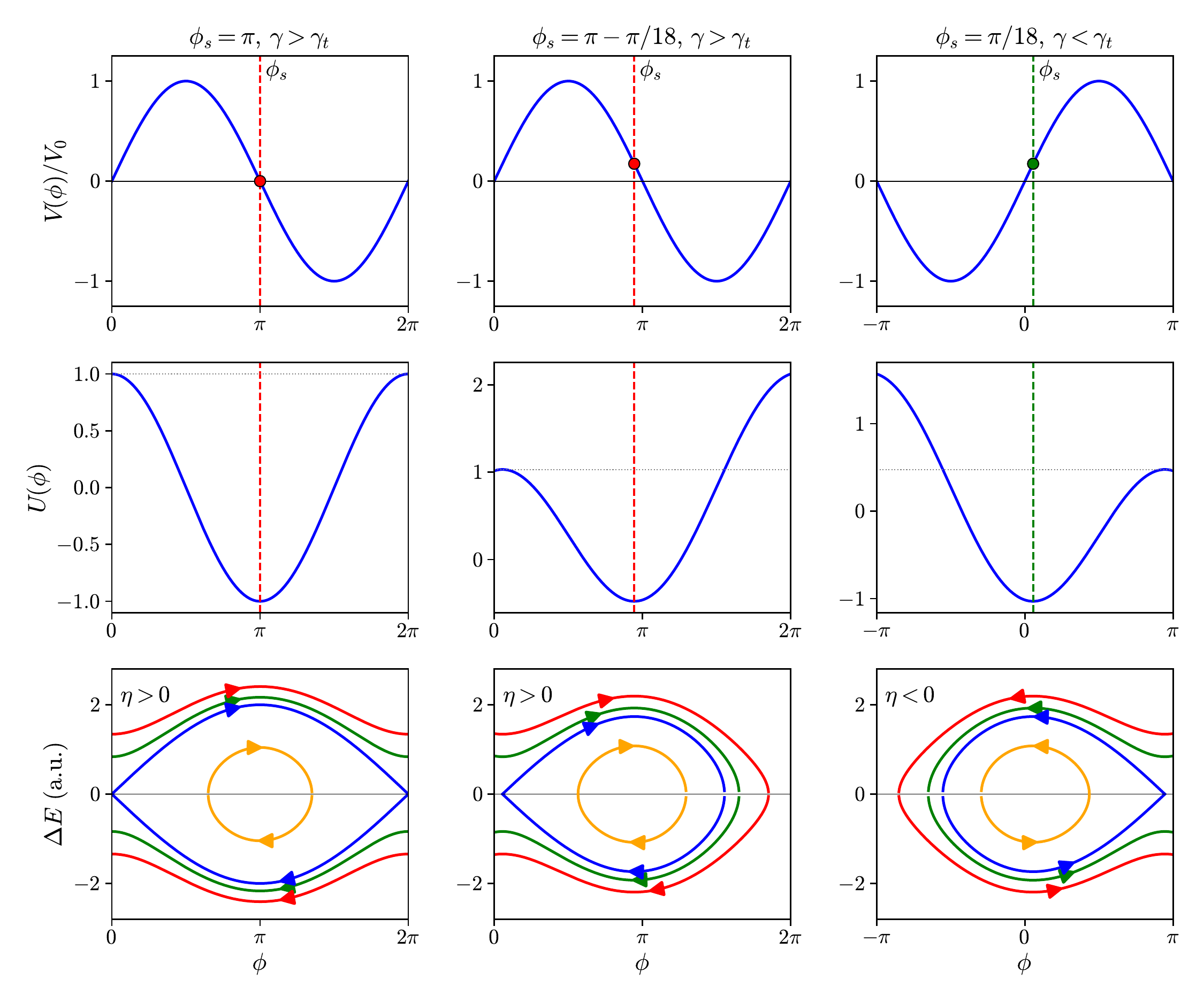}
	\caption{Top: normalised RF voltage $V(\phi)/V_0$ with the synchronous phase $\phi_s$ marked by the dashed line (colour coding as in Fig.~\ref{fig:phase_stab}: green below and red above transition). Middle: potential well $U(\phi)$, Eq.~(\ref{eq:U}); the dotted line shows the energy of synchrotron oscillations $\E$ corresponding to the separatrix. Bottom: phase space $(\phi, \Delta E)$ with the separatrix (blue) together with the bunch emittance limited by the phase trajectory (orange) and the trajectories of uncaptured particles (green and red); the arrows show the direction of particle motion. A single RF system. Left: no acceleration above transition ($\phi_s=\pi$). Middle: acceleration above transition, $\gamma>\gamma_t$, $\eta>0$, with $\phi_s =\pi-\pi/18$. Right: acceleration below transition, $\gamma<\gamma_t$, $\eta<0$, with $\phi_s = \pi/18$. Since $\dot\phi \propto \eta\Delta E$, the~direction of motion is inverted at transition crossing, while the potential well and the bucket are mirror images of each other and the synchronous phase jumps by $\pi - 2\phi_s$.}
	\label{fig:VWe}
\end{figure}

As can be seen from Eq.~(\ref{eq:U}),
the RF phase should be changed by $\pi - 2\phi_s$ at transition crossing during acceleration, see Fig.~\ref{fig:VWe}. The direction of particle motion in the longitudinal phase-space is also inverted. As follows from relation (\ref{eq:l2}), $\phi$ increases ($\dot{\phi}>0$) for particles
with $\Delta E<0$ if $\eta<0$ (below transition) and for particles with $ \Delta E>0$ if $\eta>0$ (above transition).
The extrema of the potential are defined by the condition
$U^\prime(\phi) \propto V(\phi) - V_0\sin\phi_s =0$ and thus they are the solutions of the~equation
   $ \sin\phi=\sin\phi_s$.
The phases $\phi_s +2\pi k$, $k=0,1,2,... h-1$, are the stable fixed points and $\phi_1 = \pi -\phi_s + 2\pi k$ are the unstable fixed points. The second turning point $\phi_m$ is defined by the equation
$$U(\phi_m) = U(\pi-\phi_s).$$
The trajectory (\ref{traj}) with $\E = U(\pi-\phi_s)$ is called the separatrix. It separates the bounded particle motion from the unbounded one.

The phase-space area limited by the separatrix is the RF bucket (also sometimes called the longitudinal acceptance). In the usual units of eVs it can be calculated as
\begin{equation}
    A= \frac{2}{h\omega_0} \int^{\phi_m}_{\phi_1} \Delta E(\phi) d\phi,
    \label{eq:A}
\end{equation}
where for $\Delta E$, combining Eq.~(\ref{eq:l2}) and Eq.~(\ref{traj}), we have
\begin{equation}
      \frac{h^2\omega_0^2\eta}{\beta^2 E}\frac{\Delta E_{sep} (\phi)}{h\omega_0} = \omega_{s0} \sqrt{2[U(\pi-\phi_s) -U(\phi)]}.
\end{equation}
The bucket length in phase $\Delta\phi_{bucket}$ is the distance between the unstable fixed point $\phi_1$ and the second turning point $\phi_m$ of the potential $U(\phi)$
\begin{equation}
   \Delta\phi_{bucket} = \phi_m-\phi_1 = \phi_m-(\pi-\phi_s).
\end{equation}
The relative bucket height (energy acceptance)
\begin{equation}
    \frac{\Delta E_{max}}{E} = \frac{\nu_{s0}\beta^2}{h\eta} \sqrt{2[U(\pi-\phi_s) -U(\phi_s)]},
\end{equation}
where the synchrotron tune is $\nu_{s0} =\omega_{s0}/\omega_0$.

 Without acceleration, there is no energy gain and  $\sin\phi_s=0$. In this case $U(\phi) = \cos\phi$ (for $\gamma > \gamma_t$, $\phi_s = \pi$). The bucket has the maximum area and length with $\phi_1=0$ and $\phi_m = 2\pi$.
 Then the~integral in Eq.~(\ref{eq:A}) can be easily calculated
 \[
  \int_{0}^{2\pi} \sqrt{2[U(0) - U(\phi)]} \, d\phi = 2\int_{0}^{2\pi} \sin\frac{\phi}{2} \, d\phi = 8,
 \]
so that
\[
   A = \frac{16 \beta}{\omega_{RF}}\sqrt{\frac{qEV}{2\pi h|\eta|}} \Gamma (\phi_s).
\]
 Due to acceleration, the bucket area is reduced by the factor $\Gamma (\phi_s)$,
 which is approximately~\cite{Li}
 \[
      \Gamma(\phi_s) \simeq \frac{1-|\sin\phi_s|}{1+|\sin\phi_s|}.
 \]
 Note that $A=0$ for $\phi_s=\pi/2$ or $3\pi/2$.

 \subsubsection{Bunch length and emittance}
 Particles normally fill only a part of the bucket, called a bunch. One can have up to $h=\omega_{RF}/\omega_0$ bunches in the ring. We assume that the particles are distributed around the minimum of the potential well (the~synchronous phase $\phi_s$).  Then the longitudinal emittance of the bunch is the area (in [eVs]) inside the~limiting particle trajectory with the synchrotron energy $\E_{max}$ and extreme phases $\phi_L$ and $\phi_R$  (see Fig.~\ref{fig:VWe}),
 \begin{equation}
    \varepsilon_b = \frac{2}{h\omega_{0}} \int^{\phi_R}_{\phi_L} \Delta E_b (\phi) d\phi,
    \label{eq:emit1}
  \end{equation}
where $\E_{max} = U(\phi_L)=U(\phi_R)$ and
\[
    \frac{\omega_{RF}^2\eta}{\beta^2 E}\frac{\Delta E_b (\phi)}{h\omega_0} = \omega_{s0} \sqrt{2[U(\phi_L) -U(\phi)]}.
\]
Note that for ions the units of $\varepsilon_b$ are [eVs/charge].
The full bunch length in [rad] is $\phi_R-\phi_L$ and in [s] it is $\tau = (\phi_R-\phi_L)/(h\omega_0)$.

For a non-accelerating bucket, the potential well $U(\phi) = - \cos\phi/\cos\phi_s$ is symmetric and $\phi_R-\phi_L=2\phi_b$, where $\phi_b$ is the half bunch length. In this case
\begin{equation}
    \varepsilon_b = \frac{4 E\beta^2 \omega_{s0}}{\omega^2_{RF}|\eta|} \int_0^{\phi_b} \sqrt{2(\cos\phi -\cos\phi_b)} \, d\phi.
    \label{eq:emit}
\end{equation}
For short bunches with $\phi_b \ll 1$, we can use the first two terms of the expansion $\cos\phi =1 -\phi^2/2 + ...$, so that
\begin{equation}
    \varepsilon_b^{short} \simeq  \pi \Delta E_b(0) \, \tau/2 = \pi \Delta E_b(0) \, \frac{\phi_b}{
    h\omega_0} = \frac{ \pi E\beta^2 \omega_{s0} \tau^2 }{4|\eta|}.
    \label{eq:emit_sh}
\end{equation}
However, this approximation should be applied with caution for longer bunches, since the difference between the longitudinal bunch emittances calculated using this formula and Eq.~(\ref{eq:emit1}) can be very significant, up to a factor of two, see Fig.~\ref{fig:emit_fs}(a).

Using the same conjugate variables
\{$\phi, \Delta E/(h\omega_0)$\}, the equations of longitudinal motion can also be written in Hamiltonian form
\begin{equation}
    \frac{\partial H}{\partial \phi}  = -\frac{d}{dt}  \left(\frac{\Delta E}{h\omega_0}\right)\hspace{1cm} \text{and} \hspace{1cm}
    \frac{\partial H}{\partial(\Delta E/h\omega_0)}  = \frac{d\phi}{dt}
\end{equation}
with the Hamiltonian
\begin{equation}
    H = \frac{\omega_{RF}^2\eta}{2\beta^2E} \left(\frac{\Delta E}{h\omega_0} \right)^2 - \frac{qV_0\cos \phi_s}{2\pi h}\,U(\phi).
\end{equation}
During beam acceleration, the Hamiltonian $H$ of the system 
depends on time. If the changes are slow enough (adiabatic), $H$ is considered to be quasi-static.
The parameter $\lambda$ changes adiabatically during a~period of time $T$ if $T \frac{d\lambda}{dt} \ll \lambda$~\cite{Landau}.
Applying this criterion to the synchrotron motion with the period $T_s$, we have $\frac{d T_s}{dt} \ll 1$.

The adiabatic invariant of motion, the action $J$, is directly related to the bunch emittance
\[
  J = \frac{1}{2\pi} \oint \frac{\Delta E(\phi)}{h\omega_0} d\phi = \frac{\varepsilon_b}{2\pi},
\]
therefore the longitudinal emittance is an invariant
of motion during acceleration and RF manipulations, if the parameter change is adiabatic.

 Note that even without acceleration, the average bunch energy may change due to energy loss caused by synchrotron radiation, induced voltage, electron clouds, etc. The energy lost by a particle per turn is compensated by the RF system.
As a result, the bucket becomes accelerating ($\phi_s \neq 0, \pi$) even with magnetic field $B(t)=$~const.

\subsubsection{Synchrotron frequency}
Due to the non-linearity of the RF voltage $V(\phi)$ in Eq.~(\ref{eq:phase_s}), all particles oscillate with different frequencies $\omega_s$,   which depend on the amplitude of oscillations or on $\E$. As follows from Eq.~(\ref{traj}), the period of synchrotron oscillations is
\begin{equation}
    T_s = \oint \frac{d\phi}{\dot\phi} = \frac{2}{\omega_{s0}} \int_{\phi_L}^{\phi_R} \frac{d\phi}{\sqrt{2[\E-U(\phi)]}},
    \label{Ts}
\end{equation}
where $\E = U(\phi_L)= U(\phi_R)$.

\begin{figure}[th]
	\centering
 \begin{tabular}{cc}
	\includegraphics[width = 0.48\textwidth]{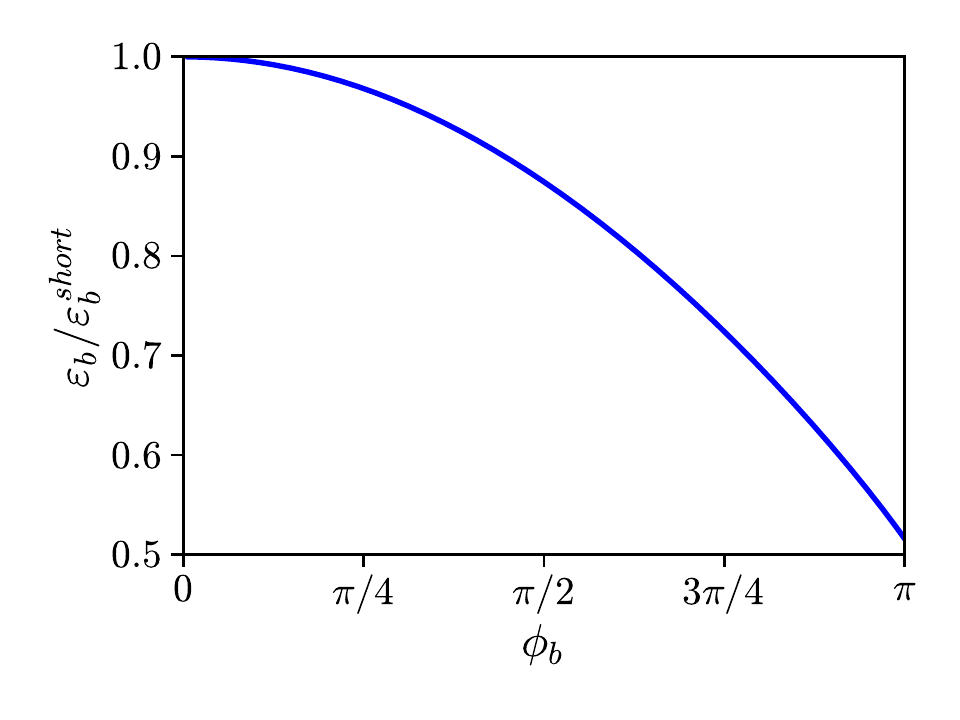} &
	\includegraphics[width = 0.48\textwidth]{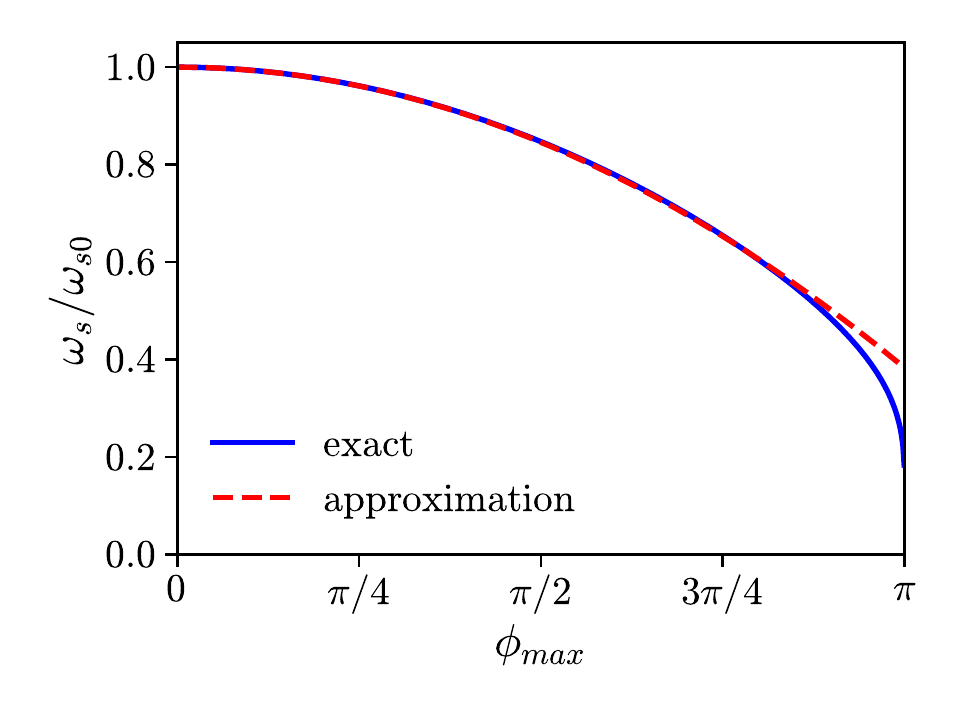}\\
	(a) & (b)
 \end{tabular}
	\caption{Comparison of the short-bunch (small-amplitude) approximation with the exact solution in a single RF system without acceleration. (a) Ratio of the exact and approximate emittances, $\varepsilon_b/\varepsilon_b^{short}$, calculated from Eqs.~(\ref{eq:emit})~and~(\ref{eq:emit_sh}) as a function of the half bunch length in radians $\phi_b$. (b) Normalised frequency of synchrotron oscillation as a function of particle amplitude $\phi_{max}$: exact Eq.~(\ref{fs_ex}), shown with blue curve, and its small-amplitude approximation (\ref{fs_app}), red dashed curve.}
	\label{fig:emit_fs}
\end{figure}

For a non-accelerating bucket in a single RF system, the synchrotron frequency of a particle with amplitude $\phi_{max}$ is
\begin{equation}
  \frac{\omega_s}{\omega_{s0}} = \frac{\pi}{2K(\sin\frac{\phi_{max}}{2})},
  \label{fs_ex}
\end{equation}
where $K(x)$ is the complete elliptic integral of the first kind. For small amplitudes
\begin{equation}
    (\omega_{s0} -\omega_{s})/\omega_{s0} \simeq   \phi_{max}^2/16,
 \label{fs_app}
\end{equation}
so that for short bunches the relative synchrotron frequency spread inside the bunch is $\phi_b^2/16$, where $\phi_b$ is the amplitude of particles at the bunch edge.
A comparison of the two solutions, the exact and the approximate, is shown in Fig.~\ref{fig:emit_fs}(b). One can see that the small-amplitude approximation in this case works well up to very large particle oscillation amplitudes.
Using the energy and phase of the~synchrotron oscillations as a new set of variables $(\E, \psi)$,
the equations of particle motion can be rewritten in the~following form
\[ \dot\E = 0, \hspace{1 cm} \dot\psi =\omega_s(\E),\]
 where $\omega_s(\E) = 2\pi/T_s(\E)$ and in the general case
\[ \psi = \text{sgn}(\eta\Delta E)\frac{\omega_s(\mathcal{E})}{\sqrt{2}\omega_{s0}} \int_{\phi_{max}}^\phi \frac{d\phi^\prime}{\sqrt{\mathcal{E} -U\left(\phi^\prime\right)}}.
\]


\section{\label{sec:double_rf}Double RF system in hadron rings}
\subsection{Parameters of a double RF system}
In hadron rings, a higher harmonic (HH) RF system is usually installed in addition to the main RF system for one of the following purposes: (1) to increase the available bucket area and reduce the peak line density; (2)  to increase the synchrotron frequency spread; (3) to modify the zero-amplitude synchrotron frequency $\omega_s(0)$; (4) to perform rf manipulations (bunch splitting, bunch rotation, controlled emittance blow-up, beam transfer, momentum slip-stacking, ...).

The HH cavities can be passive or active. In the first case, the HH voltage
is the voltage induced by the beam in the cavity
with an amplitude and a phase which are functions of the beam parameters, in particular, the beam current and bunch distribution. This option is more often used in lepton rings.
Below we consider only active (powered) rf systems.

\begin{figure}[h]
	\centering
	\includegraphics[width = \textwidth]{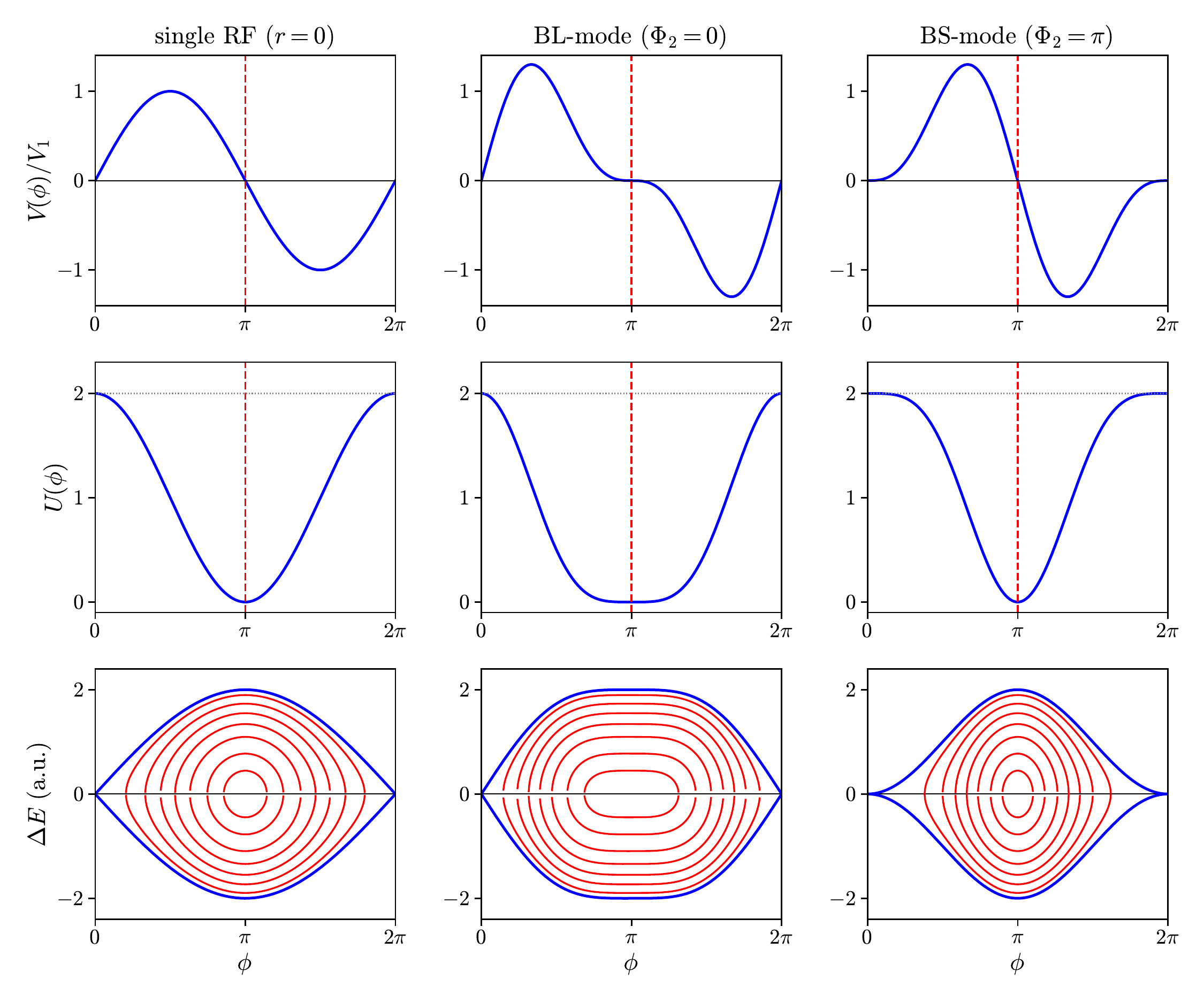}
	\caption{Examples of total RF voltage $V(\phi)$ (top), RF potential $U(\phi)$ (middle) and particle trajectories in longitudinal phase space (bottom) in single (left) and double RF system in BL-mode (middle) and BS-mode (right) for a~stationary case (no acceleration) above transition with $n=2$ and $r=1/2$. The dashed line marks the synchronous phase $\phi_s=\pi$, the dotted line the energy of synchrotron oscillations $\E$ corresponding to the separatrix (blue). The~red curves are particle trajectories with the same set of values of $\E$ in all three cases.}
	\label{fig1}
\end{figure}

The total voltage seen by the particle in a double RF system is
\begin{equation}
    V=V_1\left[\sin{\phi} +r\sin(n\phi + \Phi_n)\right],
    \label{V}
\end{equation}
where $V_1$ and $V_{n} = r V_1$ are, correspondingly, the voltage amplitudes of the main rf  and of the HH rf system with harmonic number $h_n =n h$.
Typically $V_n < V_1$ and even $V_n \leq V_1/n$.
The phase shift $\Phi_n$ defines the operation mode. The two main modes are often called bunch-lengthening (BL) and bunch-shortening (BS), due to their effect on a bunch of a given emittance for a low harmonic ratio $n$, see Fig.~\ref{fig1}.
These modes are also referred to as  ``out-of-phase'' and ``in-phase'' since the HH voltage $V_n$ is adjusted to reduce or increase, respectively, the slope of the main voltage waveform in the bunch center.

\begin{figure}[t!]
\centering
	\includegraphics[width = \textwidth]{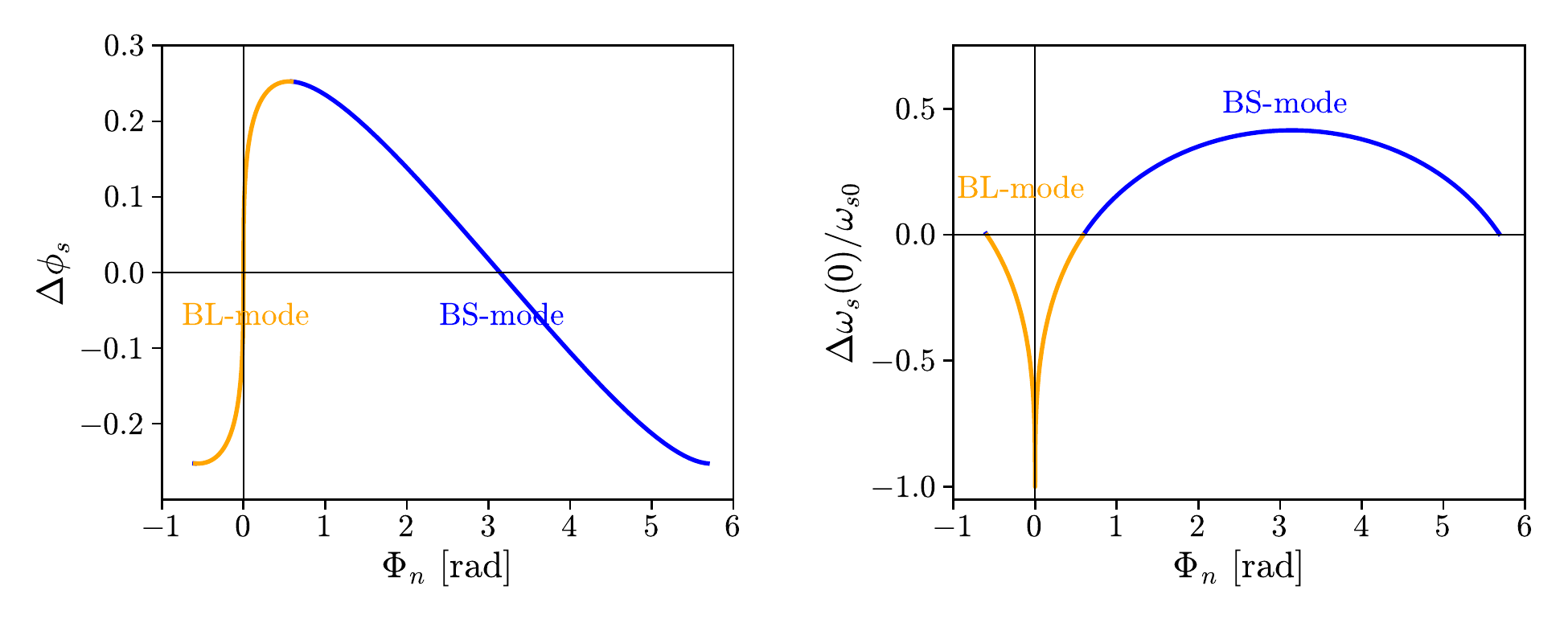}
	\caption{Synchronous phase shift $\Delta \phi_s$ (left), Eq.~(\ref{phis}), and  relative change in zero-amplitude synchrotron frequency $\Delta \omega_s(0)/\omega_{s0}$ (right), Eq.~(\ref{fs}), as functions of phase shift $\Phi_n$ in double RF system for a stationary case (no acceleration) above transition with $n=4$ and $r=1/4$. The orange parts of the curves correspond to the BL-mode, where the HH voltage reduces the slope of the total voltage in the bunch centre and $\omega_s(0) < \omega_{s0}$; note how narrow this region, $|\Phi_n| < 0.59$~rad for these parameters, is.}
	\label{fig2}
\end{figure}

For a stationary bucket, when $\sin{\phi_{s0}}=0$
(here $\phi_{s0}$ is the synchronous phase in the single, main RF system), and
for even $n$, the BL-mode is obtained with $\Phi_n= \pi$  $(0)$ below (above) transition energy.
The opposite is true for the  BS-mode or for odd values of $n$.
In the presence of the HH RF system,
the~synchronous phase is modified according to the relation
\begin{equation}
    \sin{\phi_{s0}} = \sin{\phi_{s}} + r \sin{(n\phi_{s} + \Phi_n}),
    \label{phis}
\end{equation}
see an example in Fig.~\ref{fig2} (left). At transition crossing, the phase $\Phi_n$ should be changed correspondingly.

The choice of the operational mode and its parameters is dictated by the application; some examples are presented below.

\subsection{Bucket area and bunch shape modification}

In a double RF system, in comparison to a single RF system, the bucket area is increased in BL-mode and decreased in BS-mode. The higher is $n$, the smaller is the effect. The ratio of the stationary bucket area in BL-mode with $n=2$  to that of a single RF system is a monotonic function of $r$ for $r \le 1/n$~\cite{Li}
\begin{equation}
       A(r)/A(r=0) = \left[\sqrt{1+2r} + \frac{1}{\sqrt{2r}}\ln(\sqrt{1+2r} + \sqrt{2r} )\right]/2.
\end{equation}
For $r > 1/n$, the bucket area continues to shrink in BS-mode, while in BL-mode the voltage $V(\phi)$ has additional zeros around $\phi_s$, the potential well gets two minima and two buckets start to form inside the~outer separatrix. The minimum peak line density can be obtained for $r > 1/n$, depending on the~particle distribution, see below.

In low-energy, space-charge dominated accelerators,
the BL-mode is widely used for the peak line-density  reduction, or, equivalently, increase of the bunching factor~\cite{SLAC},\cite{BF_PSB}.
Indeed, in the stationary case, the bunch line density $\lambda(U)$ is a function of the RF potential $U$ (the integral of the total RF voltage), which
in BL-mode can be made flat in the bunch center with the dependence $U(\phi) \propto \phi^4$ instead of $U(\phi) \propto \phi^2$ in the BS-mode or in a single RF system.
Such ``flat'' bunches are obtained when $V^{\prime}(\phi_s) = 0$ and $V^{\prime\prime}(\phi_s) = 0$, i.e.~\cite{Hofmann}
\begin{align}
    \cos{\phi_s} & =  - r n\cos{(n\phi_s + \Phi_n)},
    \label{1cond}\\
    \sin{\phi_s} & = - r n^2\sin{(n\phi_s + \Phi_n)}.
    \label{2cond}
\end{align}
These two conditions together with Eq.~(\ref{phis}) define $\phi_s$ and
the required $\Phi_n$ and $r$ for given $n$ and $\phi_{s0}$:
\begin{equation}
    r^2 = \frac{1}{n^2} - \frac{\sin^2{\phi_{s0}}}{n^2-1}.
    \label{cond}
\end{equation}
For $\sin{\phi_{s0}}=0$ (no acceleration) one has $r=1/n$ from (\ref{cond}) as well as  $\Phi_n = (1 -n)\phi_s + \pi$ from (\ref{1cond}) and $\phi_s=\phi_{s0}=0,\pi$ from (\ref{2cond}).


In the CERN PSB~\cite{BF_PSB}, an increase of the bunching factor $B_f$ (defined as the ratio of the mean to peak line-density) was achieved  during acceleration using the HH RF system with $n=2$, $r >1/n$  and the phase shift $\Phi_n = -n\phi_s$, which has the advantage of preserving in the double RF operation the same synchronous phase $\phi_{s}$ as in the single RF system.
For a bunch with a local elliptic energy distribution, which has the same line-density shape as the RF potential,
the minimum peak line-density is obtained for $r = 1$ and not for $r=$0.5. However, the optimum is a flat function of $r$,  and $r = 0.67$ gives only a~$3\%$ lower $B_f$.

Notice that flat bunches can also be obtained  by other methods (with or without longitudinal emittance blow-up) in a single RF system (using band-limited noise, a phase or voltage jump, etc.).
The HH rf system in BS-mode can also be used purely to reduce the bunch length~\cite{Chin}.


\subsection{Synchrotron frequency spread}

Beam stability, in general, is improved when more Landau damping, due to a larger synchrotron frequency spread, can be provided. In this application, the HH RF system is often called a ``Landau cavity''~\cite{ISR1},\cite{ISR2}.
In a double rf system, the synchrotron frequency $\omega_s$ at zero amplitude of synchrotron oscillations $\phi_{max}$ is
\begin{equation}
    \omega_s(\phi_{max}=0) = \omega_{s0} \sqrt{\frac{\cos{\phi_s} + r n \cos{(n\phi_s +\Phi_n)}}{\cos{\phi_{s0}}}}.
    \label{fs}
\end{equation}
For the same HH RF voltage (ratio $r$), a larger spread, or change in the zero-amplitude synchrotron frequency $\Delta \omega_s(0)= \omega_s(0)-\omega_{s0}$, can be obtained for a higher $n$ and in BL-mode.
The maximum change in $\omega_s(0)$ for a given value of $V_n$ is obtained for
\begin{equation}
    \Phi_n^m = - n\phi_{s0} - n\Delta\phi_s^m + \delta \Phi_n + \Phi_0,
\end{equation}
where $\delta \Phi_n$ and the corresponding synchronous phase shift $\Delta\phi_s^m$ can be found from Eqs.~(\ref{phis}),(\ref{fs}) \cite{tbohl}. The phase $\Phi_0 = 0$ or $\pi$, depending on the operation mode. In BS-mode, away from the transition crossing, $|\Delta\phi_s^m|, \; |\delta\Phi_n| \ll 1$, hence using  above transition $\Phi_n^m = - n\phi_{s0} + \pi$ during the ramp is justified. In BL-mode, small errors in $\Phi_n$ lead to a large shift in $\phi_s$, see the example in Fig.~\ref{fig2} (left). For  $\sin{\phi_{s0}}=0$, the maximum change is obtained when $\phi_s =\phi_{s0}$ and $\Phi_n =0$ or $\pi$, with the maximum $\Delta \omega_s(0)/\omega_{s0} =-1$ for $rn =1$ in BL-mode, as compared to $\Delta \omega_s(0)/\omega_{s0} =\sqrt{2}-1$ in BS-mode, see Fig.~\ref{fig2} (right). Both modes give a similar spread for $rn \ll 1$.

\begin{figure}
\centering
      \includegraphics[width = \textwidth]{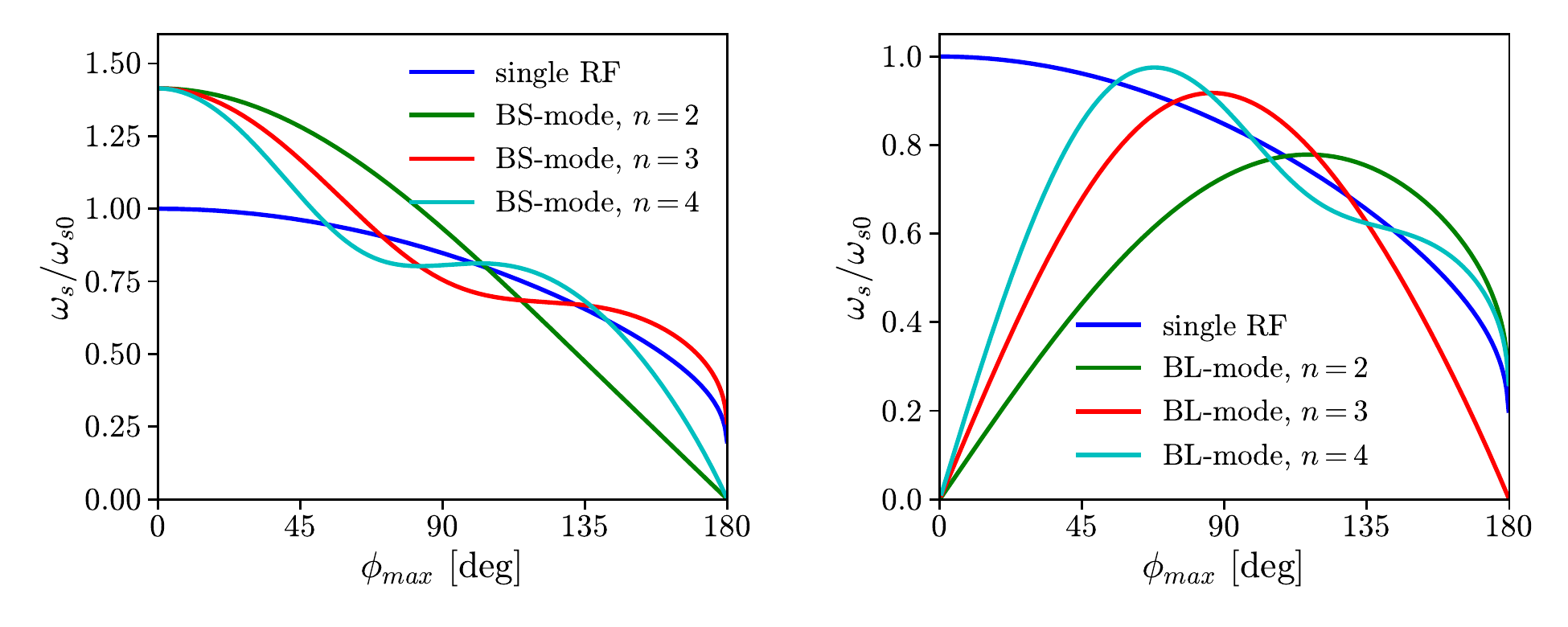}
	\caption{Synchrotron frequency $\omega_s(\phi_{max})/\omega_{s0}$ as a function of the phase amplitude $\phi_{max}$ in a stationary case in a single RF (blue curve) and a double RF system for different  $n$ and $r=1/n$ in BS-mode (left) and BL-mode (right), above transition energy, calculated from Eq.~(\ref{Ts}).}
	\label{fig3}
\end{figure}

Examples of the synchrotron frequency distribution in a double RF system for $\sin{\phi_{s0}}=0$ are shown in Fig.~\ref{fig3}. For the same $r$, in BL-mode one has a larger change in $\omega_s(0)$; however, the distribution $\omega_s(\phi_{max})$ is non-monotonic, with a region where $\omega_s^{\prime}(\phi_{max}=\phi_{cr}) = 0$.
 For $n=2$, in a stationary bucket  $\omega_s(\phi_{max})$ can be expressed analytically using the complete elliptic integral of the first kind~\cite{Li},\cite{shap}. In this case $\phi_{cr} \simeq 2.035$~rad or 117~deg.
This region inside the bunch, which is closer to the bunch center for higher $n$,  reduces the threshold of the coupled-bunch instability~\cite{BI} and may also lead to the~loss of Landau damping.
As was already noticed during the production of flat-topped bunches \cite{BF_PSB},
the bunch stability in BL-mode with $n=2$ is lost when the
bunch length approaches $\phi_{cr}=2.035$, where $\omega_s=0.7786 \omega_{s0}$. Coherent signals at this frequency were observed even for very low intensities.
Similar behavior in BL-mode was seen in other rings, including the DESY HERA and  the CERN SPS~\cite{ES}.

A reduction of the threshold of the loss of Landau damping for long bunches (close to the critical length) in BL-mode (for an inductive impedance above transition) and a significantly improved stability in BS-mode were seen in macro-particle simulations \cite{TA}, \cite{Juan}.
The threshold of the coupled-bunch instability in the CERN SPS was increased during acceleration by more than a factor of 5 in BS-mode with $\Phi_n = -n \phi_{s0}$, $n=4$ and $r < 0.2$, but it was not possible to increase the beam stability by applying the BL-mode~\cite{ES}.
Experiments confirming the advantage of BS-mode over BL-mode were also performed in the CERN PS for $n=4$~\cite{Heiko}. However, in BS-mode, long bunches approaching the~flat region in $\omega_s(\phi_{max})$, which exists for $r \gtrsim 1/n$ and $n>2$, can also become unstable due to the~enhanced radial-mode coupling \cite{IK}.
The accurate calibration and control of $\Phi_n$ during the acceleration cycle, critical for the BL-mode,
are an additional operational challenge.

\subsection{RF manipulations}
Multiple-harmonic RF systems are used in synchrotrons for various rf gymnastics (see e.g. \cite{Handbook_RG}):

(1) Bunch manipulations, such as bunch splitting, merging, coalescing, and batch compression.

(2) Controlled longitudinal emittance blow-up can be performed by the phase modulation of a HH RF system. This technique, with a high value of $n$, around 20, is used in the CERN PS,
and with $n=4$ in the CERN PSB.
Another method, using band-limited phase noise injected in the fundamental RF system, relies on the much wider synchrotron frequency spread, as compared with a single RF system, provided by a HH RF system in the BS-mode (e.g. $n=4$ in the CERN SPS).

(3) During momentum slip-stacking, the application of a 2nd-harmonic RF system can compensate for the disturbance caused by the second group of  RF cavities operating at a close rf frequency and therefore significantly reduce particle losses and longitudinal emittance blow-up~\cite{MSS}.

Transient beam loading (e.g. due to the bunch gaps) displaces bunches and modifies the phase between the two rf systems seen by different bunches in a train, creating problems for beam stability as well as for beam manipulations (for example, bunch splitting or controlled emittance blow-up~\cite{TA}).
Note also that more RF power is required for the BL-mode than for the BS-mode: for a second-harmonic RF system at 800~MHz in the LHC, estimates give more than 300 kW of rf power for 1~MV of voltage in BL-mode and less than 80~kW in the BS-mode~\cite{power}.


\begin{thebibliography}{}
\bibitem{Lee} S.~Y.~Lee, Accelerator Physics, 4th ed. (World Scientific, Singapore, 2019).
\bibitem{Ivan} I.~Karpov, T.~Argyropoulos and E.~Shaposhnikova, Thresholds for loss of Landau damping in longitudinal plane, Phys. Rev. Accel. Beams 24, 011002 (2021).
\bibitem{Karpov2024} I.~Karpov and E.~Shaposhnikova, Generalized threshold of longitudinal multibunch instability in synchrotrons, Phys. Rev. Accel. Beams 27, 074401 (2024).
\bibitem{Li} J.~Y.~Liu \emph{et al.}, Analytical solution of particle motion in a double rf system, Part. Accel. 49, 221 (1995).
\bibitem{Landau} L.~D.~Landau and E.~M.~Lifshitz, Mechanics, 3rd ed. (Pergamon Press, Oxford, 1976).
\bibitem{SLAC} R.~Gram and P.~Morton,
Advantages of a set of second harmonic rf cavities, SLAC report TN-67-30 (1967).
\bibitem{BF_PSB} J.~M.~Baillod, L.~Magnani, G.~Nassibian, F.~Pedersen and W.~Weissflog,
A second harmonic rf system with feedback-reduced gap impedance for accelerating flat-topped bunches in the CERN PS Booster,
IEEE Trans. Nucl. Sci. NS-30, 3499 (1983).
\bibitem{Hofmann} A.~Hofmann and S.~Myers,
Beam dynamics in a double rf system,
Proc. 11th Int. Conf. on High-Energy Accelerators, Geneva, Switzerland, 1980, p.~610.
\bibitem{Chin} Y.~H.~Chin,
Double rf system for bunch shortening, LBL report LBL-29622, ESG-108 (1990).
\bibitem{ISR1} P.~Bramham \emph{et al.}, ISR performance reports, runs 756, 796, 810, CERN ISR-RF/TH-AH/cr (1977).
\bibitem{ISR2} P.~Bramham, S.~Hansen, A.~Hofmann and E.~Peschardt, Longitudinal instabilities of bunched beams in the ISR, IEEE Trans. Nucl. Sci. NS-24, 1490 (1977).
\bibitem{tbohl} T.~Bohl, T.~Linnecar, E.~Shaposhnikova and J.~T\"uckmantel, Study of different operating modes of the 4th rf harmonic Landau damping system in the CERN SPS, Proc. EPAC'98, Stockholm, Sweden, 1998, p.~978.
\bibitem{shap} E.~Shaposhnikova, Bunched beam transfer matrices in single and double rf systems, CERN Report SL/94-19 (RF) (1994).
\bibitem{BI} V.~I.~Balbekov and S.~V.~Ivanov, Longitudinal instability threshold of bunched beams in proton synchrotrons, At. Energ. 62, 98 (1987) [Sov. At. Energy 62, 117 (1987)].
\bibitem{ES} E.~Shaposhnikova, T.~Bohl and T.~Linnecar, Beam transfer functions and beam stabilisation in a double rf system, Proc. PAC'05, Knoxville, TN, USA, 2005, p.~2300.
\bibitem{TA} T.~Argyropoulos, Longitudinal beam instabilities in a double rf system, PhD thesis, National Technical University of Athens, CERN-THESIS-2015-421 (2015).
\bibitem{Juan} J.~F.~Esteban M\"uller, Longitudinal intensity effects in the CERN Large Hadron Collider, PhD thesis No.~7077, EPFL, Lausanne (2016).
\bibitem{Heiko} H.~Damerau, A.~Lasheen and E.~Shaposhnikova, Higher-harmonic rf system for Landau damping in the CERN PS, Proc. IPAC'18, Vancouver, BC, Canada, 2018, p.~728.
\bibitem{IK} I.~Karpov, Longitudinal mode-coupling instabilities of proton bunches in the CERN Super Proton Synchrotron, Phys. Rev. Accel. Beams 26, 014401 (2023).
\bibitem{Handbook_RG} R.~Garoby, RF gymnastics in a synchrotron, in Handbook of Accelerator Physics and Engineering, 2nd ed., ed. A.~W.~Chao \emph{et al.} (World Scientific, Singapore, 2013).
\bibitem{MSS} J.~Eldred and R.~Zwaska,
Enhanced dynamical stability with harmonic slip stacking,
Phys. Rev. Accel. Beams 19, 104001 (2016).
\bibitem{power} T.~Roggen, P.~Baudrenghien and R.~Calaga,
A higher harmonic cavity at 800~MHz for HL-LHC, Proc. SRF2015, Whistler, BC, Canada, 2015, paper THPB017.

\end{thebibliography}
\end{document}